\documentclass[a4paper,floatfix,rmp,twocolumn,showkeys,superscriptaddress]{revtex4}
\usepackage{graphicx}
\usepackage{soul}
\usepackage{color} 
\usepackage{amsmath}
\usepackage{fancyhdr}

\begin{document}

\title{Modelling Lipid Competition Dynamics in Heterogeneous Protocell Populations}

\lhead{Shirt-Ediss et al.}
\rhead{Modelling Protocell Competition}

\providecommand{\CSL}{ICREA-Complex Systems  Lab, Institut de Biologia Evolutiva, CSIC-UPF,  Barcelona, Spain}
\providecommand{\SFI}{Santa Fe  Institute, 1399 Hyde  Park Road, Santa Fe NM 87501,  USA}  
\providecommand{\SSBIO}{Biophysics Unit (CSIC-UPV/EHU), University of The Basque Country, Spain}
\providecommand{\SSFIL}{Logic and Philosophy of Science Department, University of The Basque Country, Spain}
\providecommand{\CHEM}{Chemistry Department, University of Bari, Italy}

\author{Ben Shirt-Ediss}
\affiliation{\CSL}
\affiliation{\SSFIL}
\author{Kepa Ruiz-Mirazo}
\affiliation{\SSBIO}
\affiliation{\SSFIL}  
\author{Fabio Mavelli}
\affiliation{\CHEM}
\author{Ricard V. Sol\'e\footnote{Corresponding author: ricard.sole@upf.edu}}
\affiliation{\CSL}
\affiliation{\SFI}

\begin{abstract}

In addressing the origins of Darwinian evolution, recent experimental work has been focussed on the discovery of simple physical effects which would provide a relevant selective advantage to protocells competing with each other for a limited supply of lipid. In particular, data coming from Szostak's lab suggest that the transition from simple prebiotically plausible lipid membranes to more complex and heterogeneous ones, closer to real biomembranes, may have been driven by changes in the fluidity of the membrane and its affinity for the available amphiphilic compound, which in turn would involve changes in vesicle growth dynamics. Earlier work from the same group reported osmotically-driven competition effects, whereby swelled vesicles grow at the expense of isotonic ones. In this paper, we try to expand on these experimental studies by providing a simple mathematical model of a population of competing vesicles, studied at the level of lipid kinetics. In silico simulations of the model are able to reproduce qualitatively and often quantitatively the experimentally reported competition effects in both scenarios. We also develop a method for numerically solving the equilibrium of a population of competing model vesicles, which is quite general and applicable to different vesicle kinetics schemes.

\end{abstract}


\maketitle

%
%

\section{Introduction}\label{sec:intro}

A fundamental problem in biology involves the origins of an innovation that allowed the development of organisms in our biosphere, beyond complex chemical reaction networks: the emergence of cells \citep{MajorTransitions,LuisiBook}. Cells define a clear scale of organization and, given their spatially confined structure, they constitute efficient units where molecules can easily interact, coordinate their dynamical patterns and establish a new level of selection. However, although it is often assumed that there was a transition from some type of `less-organised' prebiotic chemistry (surely including catalytic cycles) to a cell-based living chemistry, little is yet known concerning the potential pathways that could be followed to cross it. Once in place, protocell assemblies would require available resources for their maintenance and, thus, would naturally get inserted in diverse competitive dynamics in which the main selective unit would be the whole protocellular system. In this context, aggregate-level evolution is the right scale of analysis to be considered. 

Different types of protocellular systems of diverse complexity have been studied from a theoretical standpoint \citep{varelaMaturanaUribe1974,ganti1975,dyson1985,segre2001,soleReproductionComputation2007, maciaSole2007,mavelliRuizMirazo2007,ruizMirazo2011}. In particular, by considering the coupling of a template carrying information with vesicle replication and metabolism, it has been shown that Darwinian selection is the expected outcome of competition in a protocellular world \citep{munteanu2007}. However, an early pre-Darwinian stage in the development of biological organisms was likely to be dominated by supramolecular systems disconnected from information, closer to elementary forms of metabolism and strongly constrained by the molecular diversity of the available chemical repertoire. What type of competition and cooperation processes were at work in the chemical world leading to the emergence of early protocells? Here, in that context, processes able to favour asymmetries in the chemical composition of vesicles should be expected to play a relevant role in defining the conditions under which protocellular assemblies could thrive.

\begin{figure}
\begin{center}
\includegraphics[width=7 cm]{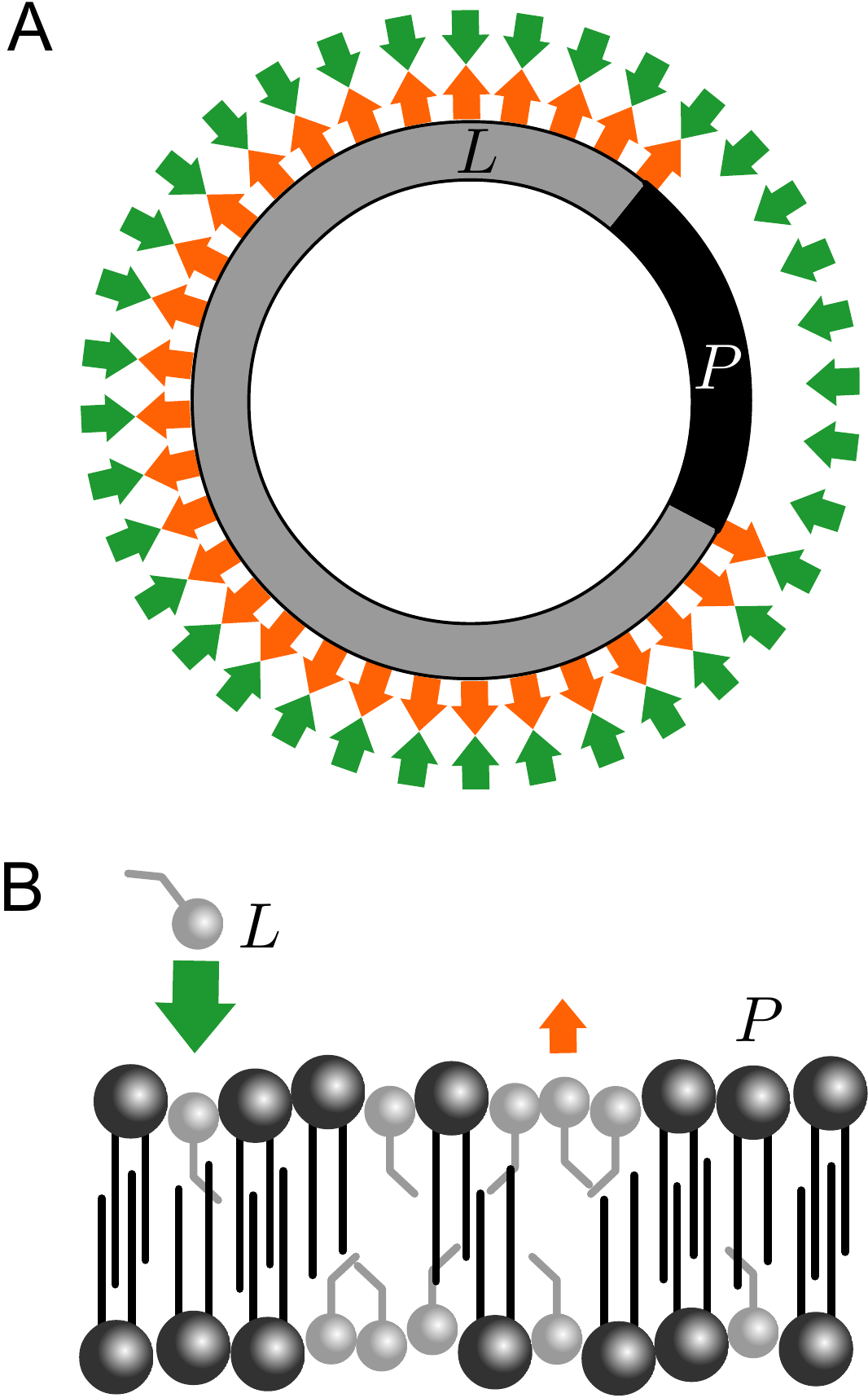}
\end{center}
\caption{
{\bf Two mechanisms of phospholipid-driven growth.} \textbf{A} \emph{Indirect effect}, whereby the presence of phospholipid in a vesicle membrane drives growth simply through a geometric asymmetry: only the lipid section of the bilayer (grey) is able to release lipid (orange arrows) whereas the whole of the bilayer surface (made of lipids and phospholipids) is able to absorb lipid monomer (green arrows). Phospholipid fraction is pictured as one continuous block to highlight the principle only. The indirect effect can be created also by non-lipid surfactant molecules (e.g. peptides) residing long enough in the membrane to increase surface absorption area. \textbf{B} \emph{Direct effect}, whereby the acyl tails of the phospholipids have high affinity for packing closer to each other and increasing bilayer order, thus making the exit of the simple lipids more difficult. The direct effect is specific to the molecular structure of phospholipids. In both cases, growth eventually stops when the phospholipid fraction in the membrane becomes diluted.
}
\label{fig:indirect_direct_mechanisms}
\end{figure}

Recent laboratory experiments have actually shown how simple physical changes made to the lipid membrane of vesicles can drive competition between those vesicles when the supply of lipid is limited. First, \citep{chenSzostak2004} reported competitive dynamics in a population of vesicles, whereby vesicles that were osmotically swollen by an encapsulated cargo of RNA (or sucrose) stole lipids from their empty osmotically relaxed counterparts by virtue of absorbing lipids more quickly. More recent experimental work has turned attention to other possible selective advantages of protocells, such as phospholipid- \citep{budinSzostak2011} and peptide- \citep{adamalaSzostak2013} driven competition amongst vesicles. Instead of membrane tension, the main factor for competition here is a different type of molecule inserted in the lipid bilayer, which changes its physical properties. In the first case, which will be the main focus of this work, fatty acid vesicles endowed with a membrane fraction of phospholipid are observed to steal lipid molecules from phospholipid-deficient neighbours, who shrink, whilst the former grow and keep their potential for division. Two basic physical mechanisms have been postulated to underlie phospholipid-driven growth, as explained in Fig. \ref{fig:indirect_direct_mechanisms} under the terms \emph{indirect} and \emph{direct} effects.

With the aim to complement experimental results, and in an attempt to better formalise and investigate competition processes at play, in this paper we develop a mathematical model of a competing population of vesicles. The model is based at the level of lipid kinetics, following the approach of \citep{ENVIRONMENT2010}. A vesicle in the population absorbs and releases lipid to and from its membrane at rates that depend on the current physical properties of that particular vesicle (such as membrane composition or extent of swelling). Using physically realistic parameters (i.e. lipid molecule sizes, vesicle aggregation numbers and CVC concentrations - see Table \ref{table:parameters}) we are able to qualitatively and often quantitatively reproduce experimental data for phospholipid-driven and osmotically driven competition.

The paper is organised as follows. The Methods section introduces the kinetic model. A mean-field analysis is performed to give insight into why we should expect phospholipid-driven competition to result from a basic version of the model kinetics, followed by the description of a fast numerical method for solving the final equilibrium state of the full model. Then, the vesicle mixing procedure is specified, in order to be able to interface the model with experimental observations. The Results section summarises how well the kinetic model is able to reproduce experimental results and observations, including also some predictions for still untested protocell competition scenarios. Finally, in the Discussion section, we consider several possible limitations of our approach and conclude the study.

%
%

\[
\,
\]
\noindent
\framebox{\begin{minipage}[t]{1\columnwidth}

\section*{Author Summary}\label{sec:author-summary}

Synthetic protocell biology is bringing forward exciting experimental results that allow us to conceive in more realistic terms how the first living organisms could have emerged and started a process of Darwinian evolution. A remarkable finding has been the capacity of lipid vesicle populations to undergo competition and selection processes without the need of nucleotide replication mechanisms. This opens a completely new research avenue to explore and characterize pre-Darwinian modes of evolution leading to the first protocellular systems. In this work, we develop a mathematical model of vesicle competition to complement ongoing experimental efforts and to also provide a reliable way to investigate scenarios or conditions that are difficult to survey in the wet lab. Our model, which is based at the level of lipid kinetics, is demonstrated to reproduce diverse reported results and is helpful in providing new insights about the molecular mechanisms underlying protocell growth and competition dynamics.

\end{minipage}}

%
%

\section{Methods}


\subsection{Theoretical Model of Vesicle Competition}

The competition model involves a set of $n$ vesicles 

\[
{\cal V} = \{  {\cal V}_1, ..., {\cal V}_n \}
\]

each one characterized by a quadruple 

\[
{\cal V}_j = (L_\mu^j, P_\mu^j, L_c^j, B_c^j)
\]

embedded in a finite volume environment $\cal E$ defined by a triple $(\Omega_e, L_e,B_e)$. Each competing vesicle consists of a unilamellar membrane of two different lipid types: a fixed number of phospholipids $P_\mu$ (e.g. di-oleoyl-phosphatidic acid, DOPA) and a variable number of simple fatty acid lipids $L_\mu$ (e.g. oleic acid, OA). The $L$ lipids in the bilayer continuously exchange with the vesicle internal water pool (also considered a well-mixed chemical domain), and $\cal E$, whereas the $P$ phospholipids are considered approximately stationary due to their comparatively slow exchange rate (a reasonable assumption, given the very low CVC values of standard phospholipids compared to other species in water). The internal water pool of each vesicle hosts $L_c$ lipid monomers and also $B_c$ buffer species, which cannot permeate the bilayer but provide osmotic stability. These buffer species are also present in $\cal E$ with constant number $B_e$.

\begin{figure*}
\begin{center}
\includegraphics[width=13 cm]{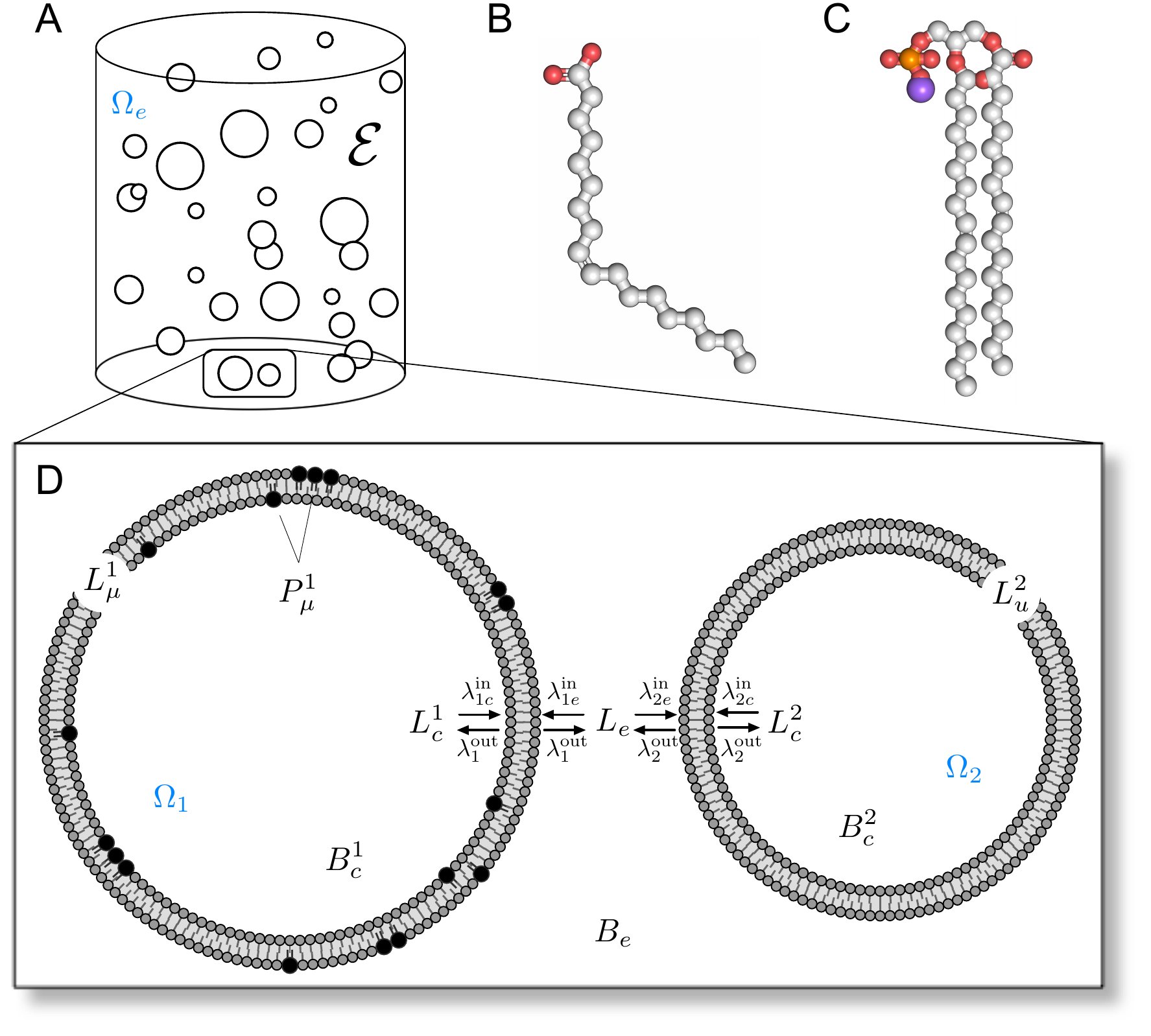}
\end{center}
\caption{
{\bf Kinetic model of vesicle competition.} \textbf{A} Our model approach considers as a starting point a population of vesicles (of generally heterogeneous sizes and membrane compositions) in a well-mixed environment. \textbf{B} Each vesicle has a membrane composed of simple single chain lipids $L$, e.g. oleic acid (OA), and \textbf{C} sometimes more complex double chain phospholipids $P$, e.g. di-oleoyl-phosphatidic acid (DOPA). \textbf{D} outlines the kinetic interactions between vesicles. Here two vesicles are displayed. Vesicle 1, on the left hand side, has a mixed membrane with approximately 10 mol \% phospholipids $P$ (black) and the remainder single chain lipids $L$ (grey). Vesicle 2 consists purely of simple lipids $L$. In the ensuing competition, phospholipid-laden vesicle 1 will grow at the expense of vesicle 2, which will shrink.}
\label{fig:schematic}
\end{figure*}

Vesicles compete with each other by consequence of uptaking/releasing simple lipids $L$ from/to $\cal E$, which is a common limited resource. The initial system of vesicles is taken to be the result of mixing different vesicle populations, and is a closed system in a non-equilibrium state. The system equilibrates to a final state following the dynamics described below, with some vesicles growing bigger in surface at the expense of others, which shrink. We ignore spatial correlations and the possibility of direct vesicle-vesicle interactions, and assume a well-mixed set of vesicles (Fig. \ref{fig:schematic}).

More precisely, each vesicle ${\cal V}_j$ is considered to release lipids to both aqueous phases (at each side of the bilayer) at the equal rate of $\lambda_j^\text{out}=k_\text{out} L_\mu^j {\textbf r}(\rho_j)$, and absorb lipids from each phase at rate $\lambda_j^\text{in}= k_\text{in} S_\mu^j [L] {\textbf u}(\Phi_j)$, where $[L]$ is the molar concentration of lipid monomer in the respective phase. The uptake and release kinetics are symmetric on each side of the bilayer, which means that the lipid monomer concentration inside and outside each vesicle will be equal $[L]_c^j = [L]_e = [L]^*$ at equilibrium. Flip-flop of the simple lipid $L$ between membrane leaflets is considered very fast with respect to its uptake and release rates, and thus the bilayer is modelled as a single oily phase.

The total number of lipids in the system $L_t$ is a conserved quantity set by the initial condition of mixing, always equal to the number of lipid monomers in the environment $L_e$, plus the number of lipids composing the vesicles. Therefore, at all times:

\begin{equation} 
 L_e + \sum_{j=1}^n \left( L_c^j + L_\mu^j \right) - L_t = 0
\label{eq:L_t-conservation}
\end{equation}

The state of the system is captured by enumerating the number of lipids in each of the aqueous pools inside the vesicles, and each of the vesicle membranes. The ODE system consists of $2n$ equations, two for each vesicle:

\begin{equation} 
\frac{dL_c^j}{dt} = k_\text{out} L_\mu^j {\textbf r}(\rho_j) - k_\text{in} S_\mu^j [L]_c^j {\textbf u}(\Phi_j)
\label{eq:L_c-ODE}
\end{equation}

\begin{equation} 
\frac{dL_\mu^j}{dt} = -2k_\text{out} L_\mu^j {\textbf r}(\rho_j) + k_\text{in} S_\mu^j ([L]_c^j+[L]_e) {\textbf u}(\Phi_j)
\label{eq:L_mu-ODE}
\end{equation}

and $L_e$ can be deduced from constraint (\ref{eq:L_t-conservation}), once all $L_c$ and $L_\mu$ have been calculated at time $t$.

Explaining the choice of lipid $L$ release kinetics, each lipid in a pure $L$ membrane is considered to have a uniform probability per unit time $k_\text{out}$ of disassociating from the membrane, and function ${\textbf r}$ modifies this probability based on the current molecular fraction of phospholipid in the membrane $\rho = \frac{P_\mu}{P_\mu + L_\mu}$. In order to account for the \emph{direct effect}, we define function $0 \le \textbf{r}(\rho) \le 1$ to be monotonically decreasing with increasing $\rho$, meaning that increasing phospholipid fraction generally decreases bilayer fluidity, slowing down the rate of $L$ release from the membrane \citep{budinSzostak2011}. In a first approximation, $\textbf{r}$ is assumed linear:

\begin{equation}
\textbf{r}(\rho)=1-d\rho
\label{eq:function-r}
\end{equation}

where parameter $0 \le d \le 1$ tunes how the lipid release rate is affected by phospholipid content ($1$ being maximally affected and $0$ being not at all).

Conversely, lipid uptake kinetics reflect that the probability of uptaking a lipid $L$ to the membrane is proportional to the density of lipid monomer in the immediate vicinity of the respective bilayer surface (i.e. the concentration of lipid in the surrounding medium), the area of surface available for absorption $S_\mu$ and function ${\textbf u}$, based on the dimensionless geometric factor $\Phi= S_{\mu} / \sqrt[3]{36\pi \Omega^{2}}$ (where $\Omega$ is vesicle aqueous volume, in the same units as $S_\mu$). We define a conditional function

\begin{equation}
\textbf{u}({\Phi})=
\begin{cases}
    exp\left(\frac{1}{\Phi}-1\right),          	& \Phi<1\\
    1,                                			& \Phi\ge1
\end{cases}\label{eq:function-u}
\end{equation}

to denote that lipid uptake is only increased when the the bilayer is stressed ($\Phi<1$). Flaccid vesicles do not have extra enhancement of lipid uptake rate. Rationale for this function stems from \citep{ENVIRONMENT2010}, to account for osmotically-driven competition between vesicles \citep{chenSzostak2004}.

To clarify some final assumptions, the vesicle surface area, referred to as $S_\mu = \frac{1}{2} (L_\mu \alpha_L + P_\mu \alpha_P)$, is the water-exposed area of the inner bilayer leaflet, or alternatively the water-exposed area of the outer bilayer leaflet of the vesicle. Membrane thickness is therefore considered negligible. Uptake and release kinetic constants $k_\text{in}$ and $k_\text{out}$ are set taking into account that spherical vesicles made purely of $L$ should be in equilibrium when the lipid monomer concentration inside and outside the vesicle is the experimentally observed CVC concentration for that amphiphilic compound (e.g. oleic acid), and also considering that $L$ uptake is orders of magnitude faster than $L$ release. For mixed membrane vesicles containing both $L$ and $P$ lipids, we assume that the lipid kinetics equations define what lipid monomer concentration inside and outside the vesicle $[L]_\text{eq}$ is necessary to keep the mixed membrane vesicle in equilibrium (however, in reality, the CVC of mixed lipid solutions is not a trivial matter \citep{cape2011}).

For the purpose of lipid competition, $\cal E$ has a fixed volume of $\Omega_e$ litres, and each vesicle ${\cal V}_j$ has, in principle, a variable volume internal water pool of $\Omega_j = \Omega_e(L_c^j+B_c^j)/(L_e+B_e)$ litres. This condition ensures that, at all times, the interior of each vesicle is isotonic with respect to $\cal E$. However, we make the simplifying assumption in this work that vesicles exist in a solution with a comparatively high buffer concentration. Thus, each vesicle has an approximately constant aqueous volume $\Omega_j \approx \Omega_e(B_c^j/B_e)$ largely determined by the number of buffer molecules it has trapped inside the internal water pool, with $L$ flux to and from the water pool having marginal osmotic effects. Model parameters are given in Table \ref{table:parameters}.


\subsection{Mean Field Approximation}

\begin{figure}
\begin{center}
\includegraphics[width=8.3cm]{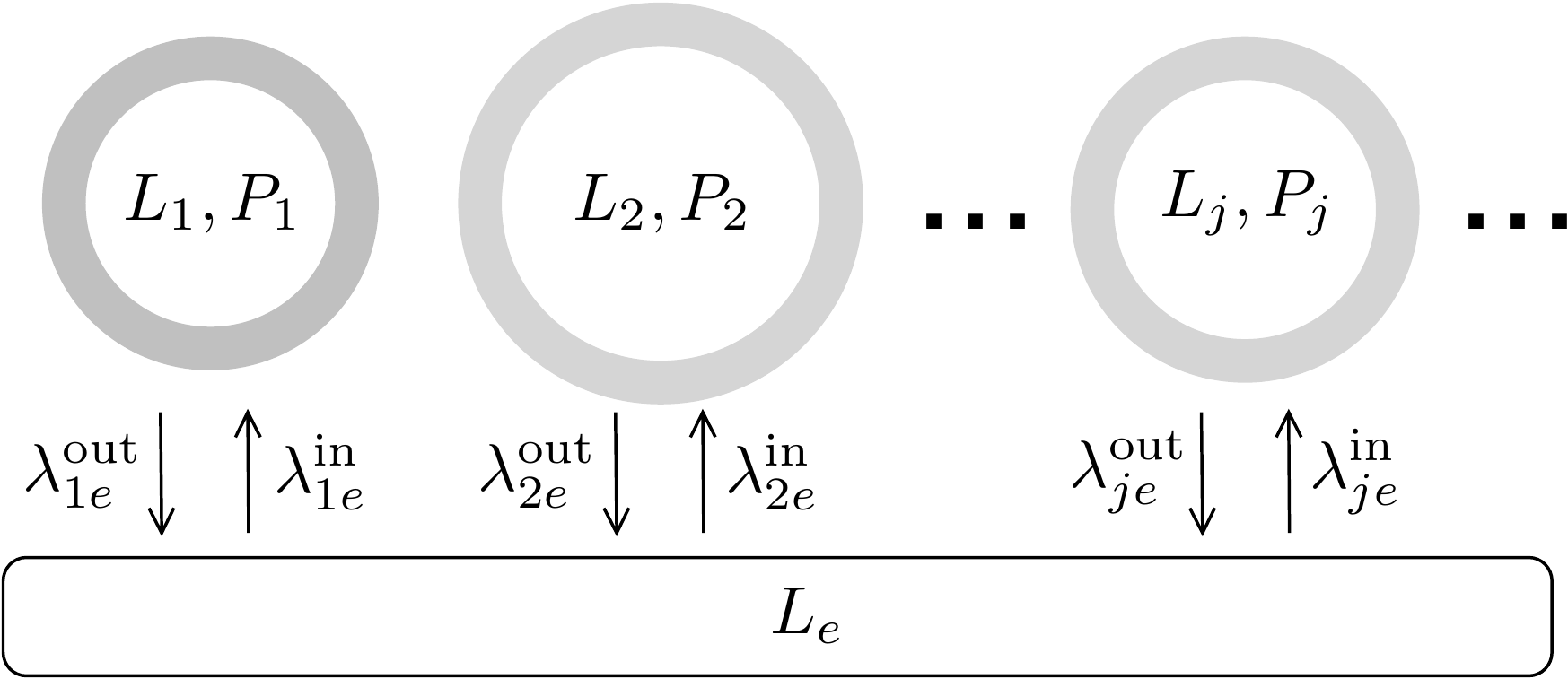}
\end{center}
\caption{
{\bf Meanfield model of vesicle population dynamics.} Considering vesicles as simplified aggregates permits some analytical treatment.}
\label{fig:meanfield_ricard}
\end{figure}

With the goal to gain intuition about why one should expect phospholipid fraction and surface growth to be correlated in the vesicle population model described, we can make a mean field approximation. This approach considers a reduced scenario where many details associated to the full model are ignored in order to keep only the logic of the problem (Fig. \ref{fig:meanfield_ricard}).

The first simplification will be to ignore the internal structure of the vesicles, describing them instead as coarse-grained `aggregates', denoted by pairs ${\cal V}_j = (L_j, P_j)$, which contain just lipids and phospholipids. This step can be considered justified on the grounds that, at equilibrium, the amount of lipid monomer residing in the vesicle water pools (which typically have tiny volumes, around 1 quintillionth of a litre) is marginal as compared to the lipid composing the vesicle membranes.  Since the internal structure or topology of the vesicles is disregarded, it actually amounts to treating them as elongated micelles or flat bilayers.

The second simplification involves reducing the lipid uptake and release equations to their most basic form, independent of membrane tension (${\textbf u}(\Phi)=1$) and independent of membrane phospholipid fraction (${\textbf r}(\rho)=1$) respectively. Thus, the ODE system reduces to $n$ simplified equations, where for each aggregate:

\begin{equation} 
\frac{dL_j}{dt} = -k_\text{out} L_j + \frac{1}{2} k_\text{in} (L_j \alpha_L + P_j \alpha_P) [L]_e
\label{eq:L-ODE-aggregate}
\end{equation}

Under these conditions, at equilibrium, the molar lipid concentration in the environment $[L]_{e}=[L]_\text{eq}$ is related to the number of lipids and phospholipids in an aggregate by the following function:

\begin{equation} 
f(L_j, P_j) = [L]_\text{eq} = \frac{2k_\text{out}}{k_\text{in}} \frac{L_j}{L_j \alpha_L + P_j \alpha_P}
\label{eq:function-f-aggregate}
\end{equation}

For a fixed number of phospholipids $P_j > 0$, the mapping $f: L_j \rightarrow [L]_\text{eq}$ can be verified to be one-to-one, meaning that each aggregate is in equilibrium at only one specific outside lipid concentration, dependent on the number of lipids $L_j$ it contains. Thus, no multiple equilibria of the population are allowed from this type of aggregate dynamics.

Now consider two arbitrarily chosen aggregates $i$ and $j$ in the population of $n$ aggregates, which are competing for lipid. Their ODEs, when written as: 
\[
\frac{dL_i}{dt} = -k_\text{out} L_i + \eta (L_i \alpha_L + P_i \alpha_P)(L_t - \sum_{m=1}^n L_m)
\]

\[
\frac{dL_j}{dt} = -k_\text{out} L_j + \eta (L_j \alpha_L + P_j \alpha_P)(L_t - \sum_{m=1}^n L_m)
\]
where $\eta = {k_\text{in}}/{2N_A \Omega_e}$, are reminiscent of the Lotka-Volterra competition equations associated to species sharing and competing for a common set of resources \citep{lotka1925book}. If we look for the equilibrium solutions of the previous system, using $dL_i/dt=dL_j/dt=0$, we obtain
\begin{equation}
\frac{L_i \alpha_L + P_i \alpha_P}{L_j \alpha_L + P_j \alpha_P} =  \frac{L_i}{L_j}
\end{equation}
which leads to the following proportionality relation at equilibrium:
\begin{equation}
L_i = \left ( \frac{P_i}{P_j} \right ) L_j
\label{eq:proportionality-at-eq-aggregate}
\end{equation}
This result immediately tells us that, for a given fraction $P_i/P_j$ the relative sizes of the two chosen vesicles are correlated. Unless $P_i=P_j$ one of the vesicles will be larger and the second smaller. For each pair $(P_i, P_j)$ with $P_i \ne P_j$ a single solution is found.

When functions ${\textbf u}$ and/or ${\textbf r}$ are not constant, unless they have a trivial form, it is generally not possible to show analytically what shape the correlation between phospholipid fraction and surface growth will take. 
However, in the next section we develop a fast numerical way to find the equilibrium configuration of the fully-fledged vesicle population model, with vesicles recovering their internal structure. As compared to numerically integrating the ODE set, the method provides the extra advantages of (i) being faster and thus scaling better for large vesicle populations and (ii) being able to calculate competition `tipping points' (i.e. critical points that mark the transition between growing and shrinking) directly.


\subsection{Fast Computation of Competition Equilibrium}

In this section we provide a general numerical approach to solving the equilibrium configuration of a possibly heterogeneous population of vesicles competing for a limited supply of lipid. These vesicles may be osmotically swelled, laden with phospholipid, or a mixture of both, and can be arbitrary in number. The method allows the lipid uptake and release functions $\textbf{u}$ and $\textbf{r}$ to take arbitrary forms, subject to some requirements detailed below.

We start by defining a function $f: L_\mu \rightarrow [L]_\text{eq}$, like (\ref{eq:function-f-aggregate}), which gives the inside/outside lipid monomer concentration $[L]_\text{eq}$ necessary to maintain a particular vesicle ${\cal V}_j$ at equilibrium, given that this vesicle has a specific number of lipids/phospholipids in the membrane, and a specific volume:

\begin{equation}
f(L_\mu^j, P_\mu^j, \Omega_j)=[L]_\text{eq}^j=\frac{2k_\text{out}}{k_\text{in}} \frac{L_\mu^j}{L_\mu^j \alpha_L + P_\mu^j \alpha_P} \frac{\textbf{r}(\rho_j)}{\textbf{u}({\Phi_j})}
\label{eq:function-f}
\end{equation}

The inverse of this function yields useful information: it is the mapping of $[L]_\text{eq}$ to the number of lipids which must exist in the membrane of a particular vesicle, in order for that vesicle to be at equilibrium.

However, due to the difficulty in isolating $L_\mu$ from the potentially non-linear functions $\textbf{u}$ and $\textbf{r}$, in most cases the inverse mapping is not possible to write in closed form. Nevertheless, if uptake and release functions $\textbf{u}$ and $\textbf{r}$ make function $f$ both (i) one-to-one and onto and (ii) continuous, then it follows that the inverse mapping is a function $f^{-1}$, which can be numerically calculated for vesicle ${\cal V}_j$ by using $f$ and binary searching for an $L_\mu$ which satisfies:

\begin{equation}
f^{-1}([L]_\text{eq}, P_\mu^j, \Omega_j) = L_\mu^j\;|\;f(L_\mu^j, P_\mu^j, \Omega_j) - [L]_\text{eq} = 0
\label{eq:function-f-inverse}
\end{equation}

using appropriate search bounds (normally: $L_\mu^\text{min}=0$, $L_\mu^\text{max}=L_t$).

Crucially, having a means to calculate $f^{-1}$ gives a way of determining the total number of lipids existing in all equilibrated vesicle membranes, given that the inside/outside lipid monomer concentration in the heterogeneous vesicle mixture is $[L]_\text{eq}$. For each $[L]_\text{eq}$, we know that each vesicle has a \emph{unique} number of membrane lipids $L_\mu$, because $f^{-1}$ is itself one-to-one. This means that a certain $[L]_\text{eq}$ can only admit one single equilibrium configuration of vesicles, not multiple equilibrium configurations, and this lack of ambiguity is a desirable property for the method. 

The lipid monomer concentration $[L]^*$ inside/outside all vesicles in this single equilibrium configuration can be found by making use of the lipid conservation principle (\ref{eq:L_t-conservation}):

\begin{multline}
[L]^* = [L]_\text{eq}\;|\;\sum_{j=1}^n f^{-1}([L]_\text{eq}, P_\mu^j, \Omega_j) + \\ [L]_\text{eq}N_A\Omega_e - L_t = 0
\label{eq:L-star}
\end{multline}

That is, at $[L]^*$, the lipid making up the membranes of all equilibrated vesicles, plus the lipid monomer inside and outside the vesicles is equal to the total lipid in the system $L_t$ set by the initial condition. Expression (\ref{eq:L-star}) can also be solved by binary search of $[L]_\text{eq}$ between appropriate bounds, normally $[L]_\text{eq}^\text{(min)}=\text{max}(f(L_\mu=0))$ over all vesicles, $[L]_\text{eq}\text{(max)}=\text{min}(f(L_\mu=L_t))$ over all vesicles. Finally, knowing $[L]^*$ allows to fully reconstruct the final sizes of all vesicles at equilibrium by substituting $[L]_\text{eq}=[L]^*$ into (\ref{eq:function-f-inverse}) for each vesicle ${\cal V}_j$.

In the equilibrated population, some vesicles will have grown larger in surface area at the expense of others which will have shrunk. When a population of vesicles has competed for lipid via phospholipid-driven competition, the `tipping point' is the critical number of membrane phospholipids $P_\mu^\text{crit}$ separating those vesicles which have lost lipid from those which have gained lipid, and is found by:

\begin{equation}
P_\mu^\text{crit} = P_\mu \;|\; f^{-1}([L]^*, P_\mu^j, \Omega_j) - \frac{2S_{\mu} -  P_\mu\alpha_P}{\alpha_L} = 0
\label{eq:Pu-critical}
\end{equation}

where $S_{\mu} = S_{\mu}^0$, again solvable by binary searching, this time in the range $0 \leq P_\mu \leq 2S_{\mu}^0 / \alpha_P$, (from a pure lipid membrane to a pure phospholipid membrane). Expression (\ref{eq:Pu-critical}) amounts to asking how many phospholipids a hypothetical vesicle would require in order not to grow in surface area when the lipid monomer concentration has stabilised at $[L]^*$. Likewise, the number of phospholipids required to achieve any arbitrary surface area growth can be found by setting $S_{\mu}$ to the value desired.

The critical phospholipid number can be stated more usefully as the critical phospholipid molecular fraction

\begin{equation}
\rho_0^\text{crit}=\frac{P_\mu^\text{crit}\alpha_L}{2S_{\mu}^0+P_\mu^\text{crit}(\alpha_L- \alpha_P)}
\label{eq:rho0-critical}
\end{equation}

a vesicle has in the initial condition, a time when all vesicles have a surface of $S_{\mu}^0$.
For osmotically-driven competition, the critical volume separating shrinking vesicles from growing vesicles is found by searching (\ref{eq:Pu-critical}) for vesicle volume instead:

\begin{equation}
\Omega^\text{crit} = \Omega \;|\; f^{-1}([L]^*, P_\mu, \Omega) - \frac{2S_{\mu} -  P_\mu\alpha_P}{\alpha_L} = 0
\label{eq:Vv-critical}
\end{equation}
where $S_{\mu} = S_{\mu}^0$. This may be alternatively stated as the critical $\Phi$ in the initial condition:

\begin{equation}
\Phi_0^\text{crit}= \frac{S_{\mu}^0}{\sqrt[3]{36\pi (\Omega^\text{crit})^{2}}}
\label{eq:phi0-critical}
\end{equation}

If no sign change results when evaluating the functions (\ref{eq:function-f-inverse} - \ref{eq:Vv-critical}) at the upper and lower search bounds, then the respective equation cannot be solved by this numerical bisection approach. Otherwise typically 30 iterations of binary search were used to converge to an accurate answer\footnote{Model vesicles described in this work either contain just fatty acid membranes, or fatty acid mixed with one other type of phospholipid. Another avenue (not explored here) would be to vary the type of phospholipid from vesicle to vesicle. In this case, competition equilibrium can still be calculated by adding two more arguments to function $f$: the first would detail the head area for the phospholipid in vesicle ${\cal V}_j$; the second would describe how this phospholipid changes bilayer fluidity and alters the simple lipid $L$ release rate (e.g. parameter $d$ to function \textbf{r} could be supplied).}.


\subsection{Modelling Vesicle Mixing}

In order to interface our theoretical model with experimentally reported results, it is necessary to define an acceptable procedure for mixing two competing vesicle populations. The basic mixing procedure outlined in this section establishes the boundary conditions for competition, which are (i) the number of vesicles present, (ii) their respective compositions, (iii) the environment volume $\Omega_e$ and (iv) the total amount of lipid $L$ in the system ($L_t$). Below, turning more specific to match experimental scenarios, simple lipids $L$ are considered to be OA, and phospholipids $P$ are considered to be DOPA.

\subsubsection*{Competition Volume}

The equilibrium finding method outlined in the `Fast Computation of Competition Equilibrium' section above requires summing over a finite number of vesicles. Likewise, dynamic simulations of the model require a finite ODE set. However, vesicle populations in a real laboratory experiment will typically have millions of vesicles competing for lipid. In our modelling approach, it is therefore necessary to consider a small volume `patch' of each of the solutions being mixed. Each patch volume is large enough to contain enough vesicles so as to be \emph{representative} of the vesicle density in the solution it pertains to, but no so many vesicles that numerical solution becomes infeasibly slow. 

A patch volume $\Omega_p = \Omega_\text{stoi}$ litres (Table \ref{table:parameters}) was utilised for stoichiometric calculations using the equilibrium finding method, which translates into around 2000 vesicles being involved in 1:1 mixing. Full dynamic simulation of the model with the Gillespie Direct SSA algorithm forced a yet smaller patch volume $\Omega_{p} = \Omega_\text{dyn}$ litres to be used (to give $\tau$ jumps which were not too small), translating into around 40 vesicles being involved in 1:1 mixing.

\subsubsection*{Mixing for Phospholipid-Driven Competition}

In order to mix a fixed population of DOPA:OA vesicles which have molecular fraction $\rho$ of DOPA in their membranes, in ratio $R$ with a variable population of pure OA vesicles, we assume the following basic steps.

Firstly, a suspension of OA lipid monomers in concentration $RC_{0}$ molar (assuming $RC_{0} \gg \text{CVC}$ for oleic acid) is extruded (possibly multiple times) through 100nm diameter pores. This leads to a more homogeneous population of 100nm diameter pure OA unilamellar vesicles. Each vesicle is assumed spherical ($\Phi=1$) with aqueous volume $\Omega^0$. The molar concentration of OA vesicles in the extruded suspension is approximately

\begin{equation}
C_\text{ves}^\text{OA}=\frac{RC_{0}}{N^\text{OA}}
\label{eq:cvesOA}
\end{equation}

where $N^\text{OA}$ is called the `aggregation number', equal to the total number of lipids forming a vesicle bilayer (in this case, just OA lipids). The lipid monomer concentration in the aqueous solution inside/outside the vesicles is $[L]_\text{eq}^\text{OA}$, the CVC value, maintaining them at equilibrium.

Secondly, a mixed suspension containing both OA lipid monomers (in molar concentration $C_{0}$) and DOPA phospholipids (in molar concentration $gC_{0}$, where $g = \frac{\rho}{1-\rho}$) is extruded through
100nm diameter pores. This, similarly, leads to a population of 100nm diameter unilamellar DOPA:OA vesicles. Again each vesicle is assumed spherical ($\Phi=1$) with aqueous volume $\Omega^0$, but now part of the bilayer consists of DOPA phospholipid in molecular fraction $\rho$. The molar concentration of DOPA:OA vesicles in the extruded suspension is approximately

\begin{equation}
C_\text{ves}^\text{DOPA:OA}=\frac{C_{0}(1+g)}{N^\text{DOPA:OA}}
\label{eq:cvesDOPAOA}
\end{equation}

where the aggregation number $N^\text{DOPA:OA}$ is now calculated as the sum of both the OA lipids and DOPA phospholipids making up each closed bilayer. In turn, the OA lipid monomer concentration inside/outside the vesicles is $[L]_\text{eq}^\text{DOPA:OA}$, the CVC value for the model DOPA:OA vesicles, maintaining them at equilibrium.

Competition starts ($t=0$) when the extruded vesicle solutions above are mixed. We mix a volume $\Omega_p$ of each solution, creating a new mixed system of volume $\Omega_e = 2\Omega_p$, containing DOPA:OA vesicles in number $N_A \Omega_p C_\text{ves}^\text{DOPA:OA}$ and OA vesicles in number $N_A \Omega_p C_\text{ves}^\text{OA}$. The initial lipid monomer concentration in the environment becomes $\frac{1}{2}([L]_\text{eq}^\text{DOPA:OA}+[L]_\text{eq}^\text{OA})$. Throughout mixing, and during competition, buffer concentration is constant at $[B]$ in all solutions, at a value high enough for vesicles to maintain approximately constant volume $\Omega^0$.

Modelling the opposite scenario, namely a fixed population of pure OA vesicles mixed with a variable population of DOPA:OA vesicles, just requires switching the $R$ multiplier from (\ref{eq:cvesOA}) to (\ref{eq:cvesDOPAOA}).

\subsubsection*{Mixing for Osmotically-Driven Competition}

When a fixed population of isotonic OA vesicles is to be mixed in ratio $R$ with a variable population of swelled OA vesicles, again two extruded vesicle suspensions are prepared. The first is prepared in buffer at molar concentration $[B]$ and extruded through 100nm diameter pores, leading to unilamellar OA vesicles at $\Phi=1$ in molar concentration 

\begin{equation}
C_\text{ves}^\text{isotonic}=\frac{C_{0}}{N^\text{OA}}
\end{equation}

The second suspension is prepared in a solution which contains an additional membrane impermeable (or slowly permeating) solute, such as sucrose, mixed with the buffer, increasing the overall molar concentration of osmotically active species to $[B]_0 \ge [B] + 0.7$. This suspension is made of unilamellar OA vesicles at $\Phi=1$ in molar concentration $RC_\text{ves}^\text{isotonic}$, and each vesicle encapsulates buffer at concentration $[B]_0$.

The buffer concentration outside the vesicles in the second suspension is then reduced to $[B]$, making the external solution hypotonic with respect to the vesicle interiors. The vesicles swell to maximum size, and then transiently break, allowing for the escape of buffer molecules in excess. They later reseal with a residual buffer gradient of $[B]_\Delta^\text{max}=0.16$M across the membrane, corresponding to a maximum osmotic pressure of 4 atm \citep{chenSzostak2004}. In our model, each vesicle is therefore assumed to swell to volume $\Omega = \Omega^0(1 + (0.16/[B]))$, which remains constant for the duration of competition.

The decrease of the environmental buffer concentration is considered to take place at the same instant of mixing with the initial isotonic population. This defines the initial condition ($t=0$) when competition starts. The mixed overall volume is $\Omega_e = 2\Omega_p$ where isotonic vesicles number $N_A \Omega_p C_\text{ves}^\text{isotonic}$, and the swelled vesicles number an $R$ multiple of this. The lipid monomer concentration in this new, larger environment is initially $[L]_\text{eq}^\text{OA}$.

\subsubsection*{Vesicle Breakage}

During competition, to a first approximation, we assume that all of the original vesicles remain intact, with none breaking apart through excessive osmotic stress. By using the Morse equation for osmotic pressure and data supplied in \citep[Supplementary Material]{chenSzostak2004}, we were able to calculate an approximate burst tolerance $\epsilon \approx 0.21$ for our model pure oleate vesicles, where these vesicles burst through excessive osmotic pressure when $\Phi < 1 - \epsilon$. Pure OA vesicles reached a minimum of $\Phi=0.77$ in our phospholipid-driven competition simulations, and a minimum of $\Phi=0.70$ in our osmotic-driven competition simulations reported in Fig. \ref{fig:experimental_comparison}. These values do not overly exceed the burst tolerance.

\subsubsection*{Control Experiments: Mixing With Buffer}

Mixing a vesicle population with a buffer solution is modelled as doubling the current system volume and diluting the initial vesicle density to one half. In this case, we assume that the buffer solution contains no vesicles, but free lipid monomer at concentration just below the CVC of oleic acid\footnote{General note: The above procedures define a `concentration approach' to mixing, where two equal volumes are mixed, and the number of vesicles in the variable population is controlled by increasing or decreasing vesicle concentration. Another approach to mixing would be the `volume approach' whereby the variable population has a fixed vesicle concentration, but instead a variable volume which controls the number of vesicles present. Volume mixing was found to produce nearly equivalent outcomes, so only results following the concentration mixing procedure are here reported.}.

%
%

\section{Results}

\subsection*{Two Competing Populations: Comparison with Experimental Results}

\begin{figure*}[ht]
\begin{center}
\includegraphics[width=17.34cm]{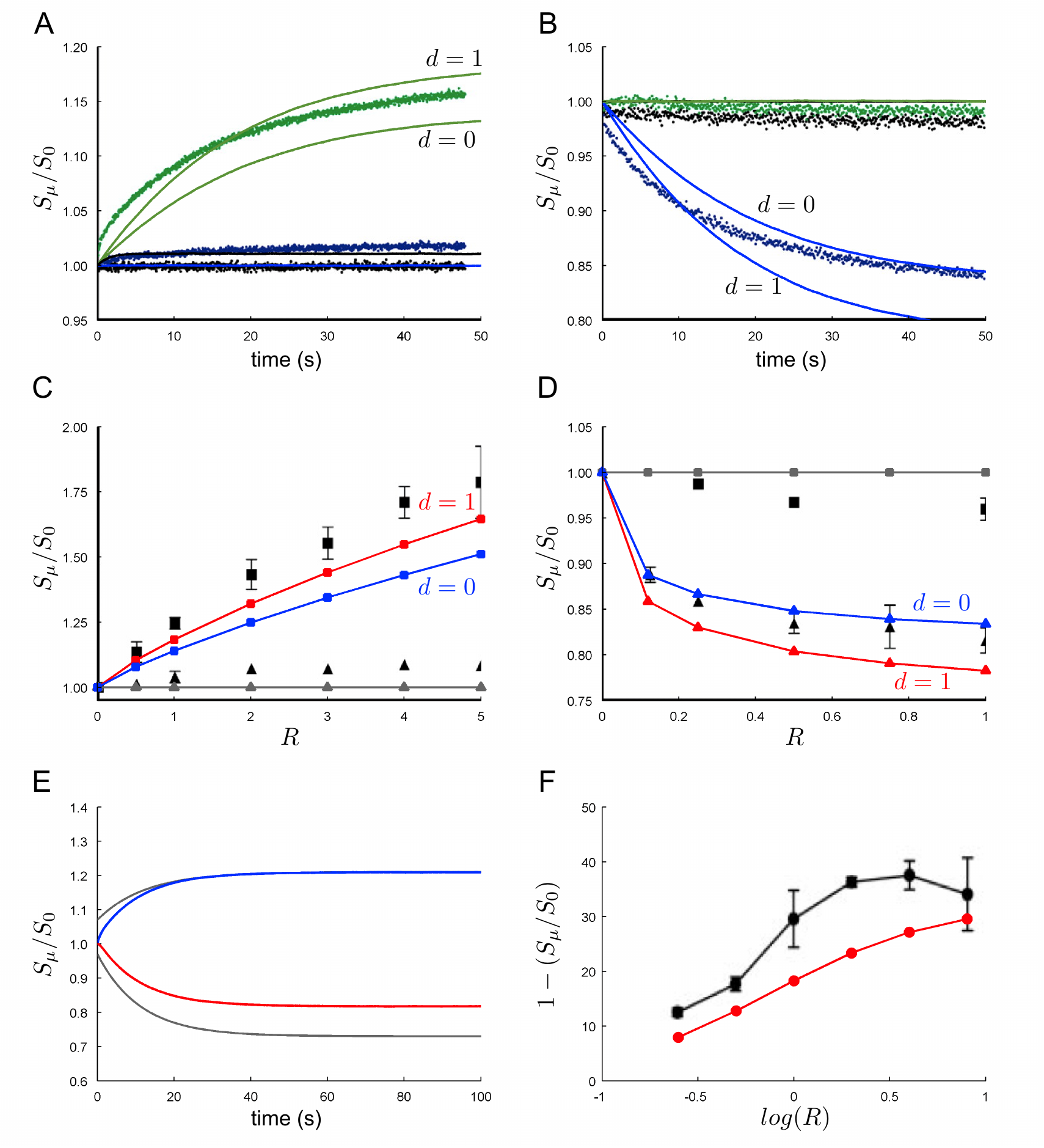}
\end{center}
\caption{
{\bf Comparison between kinetic model predictions and experimental results.} Top plots show \emph{dynamics} of phospholipid-driven competition. \textbf{A} Surface growth of DOPA:OA vesicles over time (green lines) and \textbf{B} surface shrinkage of OA vesicles over time (blue lines), when a population of DOPA:OA ($\rho=0.1$) vesicles are mixed 1:1 with pure OA vesicles (following our mixing procedure with $\Omega_{p}=\Omega_\text{dyn}$, 25 DOPA:OA are mixed with 20 OA). Coloured dots in figure backgrounds reproduce original experimental results from \cite[Figs 1A, 1B therein]{budinSzostak2011} respectively. Middle plots show \emph{vesicle stoichiometry effects} in phospholipid-driven competition. \textbf{C} Continued surface growth of DOPA:OA population as more OA vesicles added and \textbf{D} plateau in surface shrinkage of OA vesicles as more DOPA:OA vesicles added. Black markers in figure backgrounds reproduce experimental results from \cite[Figs 1C, 1D therein]{budinSzostak2011} respectively. Bottom plots show osmotically-driven competition results. \textbf{E} Growth dynamics of swelled OA vesicles (blue line) and shrinkage of isotonic vesicles (red line) compared against experimental best-fit exponential decay curves (grey lines) from \cite[Figs 1D, 1B therein]{chenSzostak2004} respectively. \textbf{F} Stoichiometry effects in osmotically-driven competition: shrinkage of OA vesicle surface reaches a plateau as more swelled vesicles are added. Black markers in figure background reproduce experimental results from \cite[Fig. 2A therein]{chenSzostak2004}. See text for discussion.}
\label{fig:experimental_comparison}
\end{figure*}

Figure \ref{fig:experimental_comparison} compares predictions made by our kinetic model against experimentally reported surface growth of vesicles in phospholipid-driven \cite{budinSzostak2011} and osmotically-driven \cite{chenSzostak2004} competition.

%

\begin{figure*}[!ht]
\begin{center}
\includegraphics[width=17.34cm]{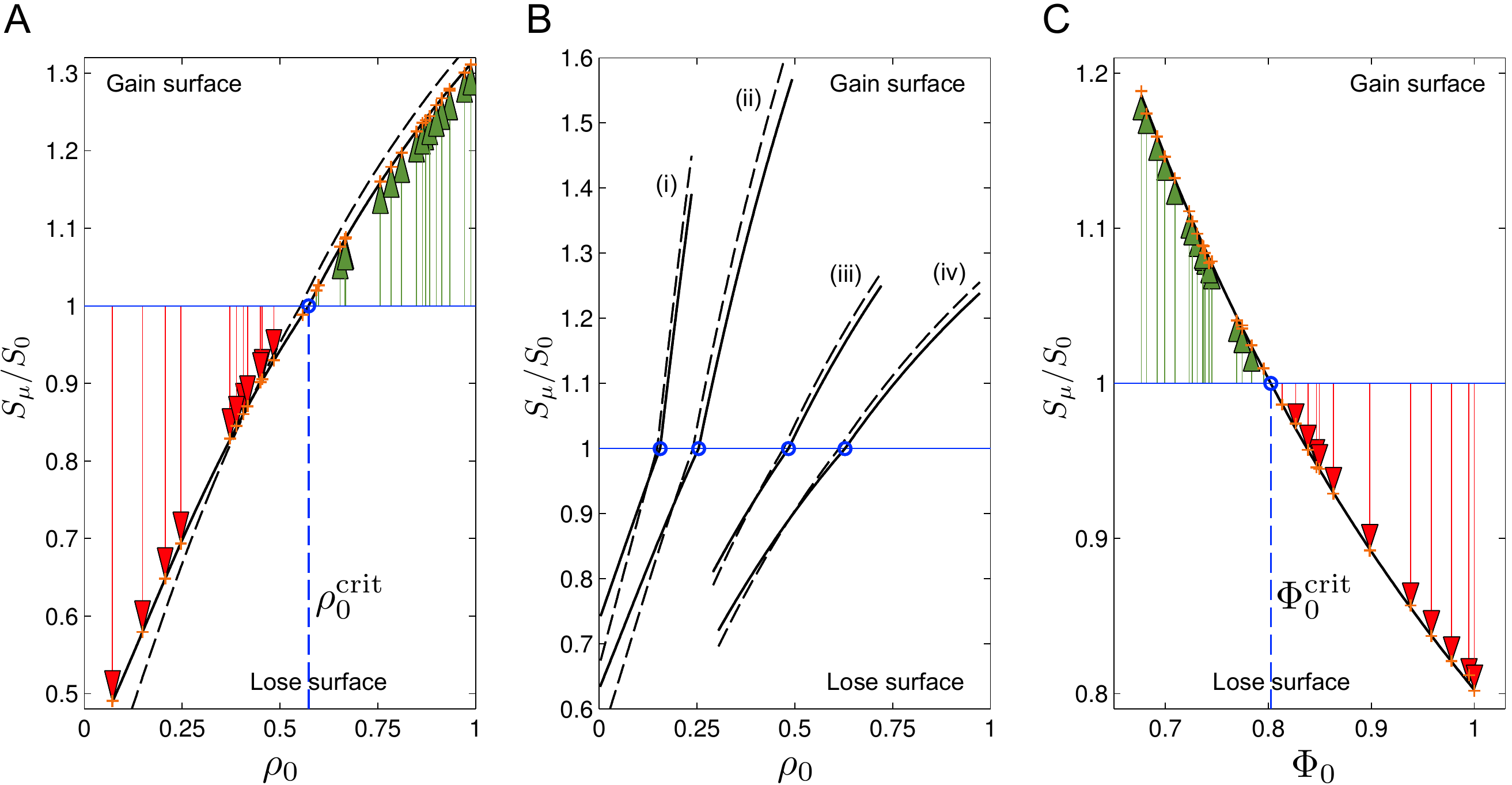}
\end{center}
\caption{
{\bf Lipid competition tipping points.} \textbf{A} Phospholipid competition between 30 model phospholipid-laden vesicles, each with DOPA fraction randomly assigned over the uniform interval $0 < \rho_0 < 1$. Depending on its initial DOPA fraction, each vesicle starts at a point on the horizontal blue line, and grows (green arrows) or shrinks (red arrows) to a point on the black line. The form of the black line is \emph{specific} to this particular competing population, and is computed by (\ref{eq:L-star}). Orange crosses show agreement with equilibrated stochastic simulation of the model, validating the fast computation of competition equilibrium method. Competition `tipping point' is shown by blue circle: any vesicle with $\rho_0^\text{crit} > 0.573$ gains lipids from its competitors. \textbf{B} Phospholipid competition in four different populations of 30 model vesicles, with DOPA fraction randomly assigned over uniform intervals (i) $0 < \rho_0 < 0.25$, (ii) $0 < \rho_0 < 0.5$, (iii) $0.25 < \rho_0 < 0.75$ and (iv) $0.3 < \rho_0 < 1.0$, demonstrating the context-dependence of competition. \textbf{C} Osmotic competition between 30 model oleate vesicles each swelled by extra internal sucrose, randomly assigned over the uniform interval $0 \le [B]_\Delta < 0.16$ molar. Any vesicle starting at tension state $\Phi_0^\text{crit} > 0.802$ gains lipids from its competitors. See text for full discussion.
}
\label{fig:tipping_points}
\end{figure*}

Top figures \ref{fig:experimental_comparison}A and \ref{fig:experimental_comparison}B show the \emph{dynamics} of surface area change in phospholipid-driven competition. Figure \ref{fig:experimental_comparison}A details, in real time, the relative surface area of a tracked population of DOPA:OA ($\rho=0.1$) vesicles, when this population is mixed $1:1$ with (i) pure OA vesicles (green lines), (ii) DOPA:OA ($\rho=0.1$) vesicles (blue lines) or (iii) buffer (black lines). In Fig. \ref{fig:experimental_comparison}B, the tracked population is instead pure OA vesicles, which are mixed $1:1$ with the same three options outlined above.

Stochastic simulation of our lipid kinetics model \cite{g1976} correctly predicts that when mixed $1:1$, DOPA:OA vesicles steal lipid and grow (green lines, \ref{fig:experimental_comparison}A) at the expense of the pure OA vesicles, which shrink (blue lines, \ref{fig:experimental_comparison}B). In this case, there is also fairly good quantitative agreement with the experimentally observed time courses, with the \emph{indirect effect} alone ($d=0$ lines) accounting for most of the surface area change in our model. For the other cases, the kinetic model correctly predicts approximately no surface area change (no competition) when similar populations are mixed, or when a population is mixed with buffer.

Middle figures \ref{fig:experimental_comparison}C and \ref{fig:experimental_comparison}D show phospholipid-driven competition from a different angle: that of \emph{vesicle stoichiometry}. Stoichiometry explores the final equilibrium size of vesicles in a tracked population, when this population is mixed with a different population containing approximately $R$ times as many vesicles. In this approach, the trend of final equilibrium surface area size versus mixing ratio is explored, rather than the dynamics on the way to equilibrium. Figure \ref{fig:experimental_comparison}C details final surface area of a tracked population of DOPA:OA ($\rho=0.1$) vesicles, when this population is mixed $1:R$ with a population of pure OA vesicles. Figure \ref{fig:experimental_comparison}D details the opposite scenario, whereby the tracked population is OA vesicles, mixed 1:R with DOPA:OA vesicles. The $R=1$ cases in Figs \ref{fig:experimental_comparison}C and \ref{fig:experimental_comparison}D correspond to the surface sizes reached in the limit of time in Figs \ref{fig:experimental_comparison}A and \ref{fig:experimental_comparison}B respectively.

Calculating competition equilibrium by means of the fast computation approach outlined in the Methods section, we were able to verify that our model exhibits continual growth of DOPA:OA ($\rho=0.1$) vesicles as more OA vesicles are added (Fig. \ref{fig:experimental_comparison}C). In the opposite scenario, we also verified that the model shows a plateau in the shrinkage of pure OA vesicles as more DOPA:OA ($\rho=0.1$) vesicles are added (Fig. \ref{fig:experimental_comparison}D). In both cases, the \emph{indirect effect} ($d=0$ lines) alone drives the majority of the surface size change, with the \emph{direct effect} then `tuning' the fit to experimental outcomes.

Importantly, the general outcome of phospholipid-driven competition in our model is for vesicles stealing lipid to grow in surface and finish at high $\Phi > 1$ values (excess surface, flaccid), and for vesicles losing lipid to suffer reduced surface, finishing at $\Phi < 1$ values (osmotically tense, spherical). This is observed experimentally, and indeed provides the basis for the conjecture that phospholipid-laden vesicles are more likely to divide spontaneously when gentle external shearing forces are applied \cite[p5250]{budinSzostak2011}.

Moving to osmotically-driven competition, Fig. \ref{fig:experimental_comparison}E shows stochastic simulation of a swelled population of vesicles competing with an initially isotonic (non-swelled) population. Simulation outcomes match quite well the experimental best-fit time courses, in particular for the growth of the swelled vesicles (not so accurately for the shrinkage of the non-swelled vesicles). In any case, it must be noted that the original experimental data has considerable variance. Then, Fig. \ref{fig:experimental_comparison}F shows that the kinetics model qualitatively reproduces the stoichiometric observation whereby adding more swelled vesicles to a population of initially non-swelled vesicles will cause the shrinkage of the non-swelled vesicles to plateau, rather than to continue (note the logarithmic scale of Fig. \ref{fig:experimental_comparison}F).

As with phospholipid-driven competition, in our model, the pure OA vesicles involved in osmotically-driven competition can finish at a variety of surface sizes. However, unlike phospholipid-driven competition, all OA vesicles will finish with the same $\Phi < 1$ value (equal osmotic stress). Here, surface changes do not translate to final differences in $\Phi$, partly because the vesicles start with different aqueous volumes. This residual osmotic swelling is also observed experimentally in vesicles stealing lipid through osmotically-driven competition. In fact, it stands as the main criticism of the osmotically-driven competition scenario: swelled vesicles have to overcome a stronger energetic barrier in order to divide, making this an improbable route to spontaneous vesicle division \cite{adamalaSzostak2013}.

Our kinetic model can also be used to make predictions or to find competition `tipping points' in the more general scenario where completely heterogeneous populations of phospholipid-laden and/or osmotically swollen vesicles compete for lipid (Figs \ref{fig:tipping_points} and \ref{fig:competition3d}), even if some of these experiments have not been realised in the lab yet.

\subsection*{Competition Tipping Points in Diverse Populations}
	
Figure \ref{fig:tipping_points}A shows that within a population of phospholipid-laden vesicles, where each vesicle has a randomly assigned phospholipid fraction in the membrane between 0 and 100\%, the critical DOPA fraction needed for growth (tipping point), in this case, is just over 57\%. Figure \ref{fig:tipping_points}B compares different heterogeneous populations competing for phospholipid, and reveals an important observation: \emph{competition is always context dependent}. That is to say, a certain amount of membrane phospholipid does not guarantee a certain final surface area. Rather, final surface depends on the boundary conditions of the competition event (that is, the parameters influencing the solution of (\ref{eq:L-star})), which includes the number and composition of competitor vesicles present\footnote{The initial lipid $L$ content of each individual vesicle is not explicitly part of these boundary conditions. In fact, in our model, when total lipid $L_t$ is fixed, initial vesicle surface sizes have no effect on the final equilibrium of the system, only on the transient dynamics leading there.}. For example, population (i) in Fig. \ref{fig:tipping_points}B has vesicles with low DOPA fraction as compared to vesicles in population (iv), yet in some cases, the vesicles in the former population have larger final surface growth than vesicles in the latter. This concurs with the experimental observation that even small differences in phospholipid content can drive growth \cite[p5251]{budinSzostak2011}.

The dotted black lines in Figs \ref{fig:tipping_points}A and \ref{fig:tipping_points}B are the same competition events run when the \emph{direct effect} is present, and maximally enabled ($d=1$). The direct effect makes the competition tipping point slightly lower, but no general statement can be made about the extent to which it affects vesicle growth, for this again depends on the specifics of the competition event. For example, the direct effect has marginal influence on vesicle growth trends in the population shown in Fig. \ref{fig:tipping_points}B (iii), but is more relevant in population (ii).

Figure \ref{fig:tipping_points}C shows that in a heterogeneous population where pure OA model vesicles are swelled with residual buffer up to 0.16M, vesicles with low initial $\Phi$ values steal lipid from those with higher (less swelled) $\Phi$ values, with the tipping point between growing and shrinking at $\Phi_0^\text{crit}=0.8$. As a last remark, orange crosses marked on Figs \ref{fig:tipping_points}A and \ref{fig:tipping_points}C show that full stochastic simulations of the model (run all the way to equilibrium) agree with and thus validate the fast computation of competition equilibrium method.

\subsection*{Theoretical Predictions Beyond Current Experimental Results}

\begin{figure*}[ht]
\begin{center}
\includegraphics[width=17.34cm]{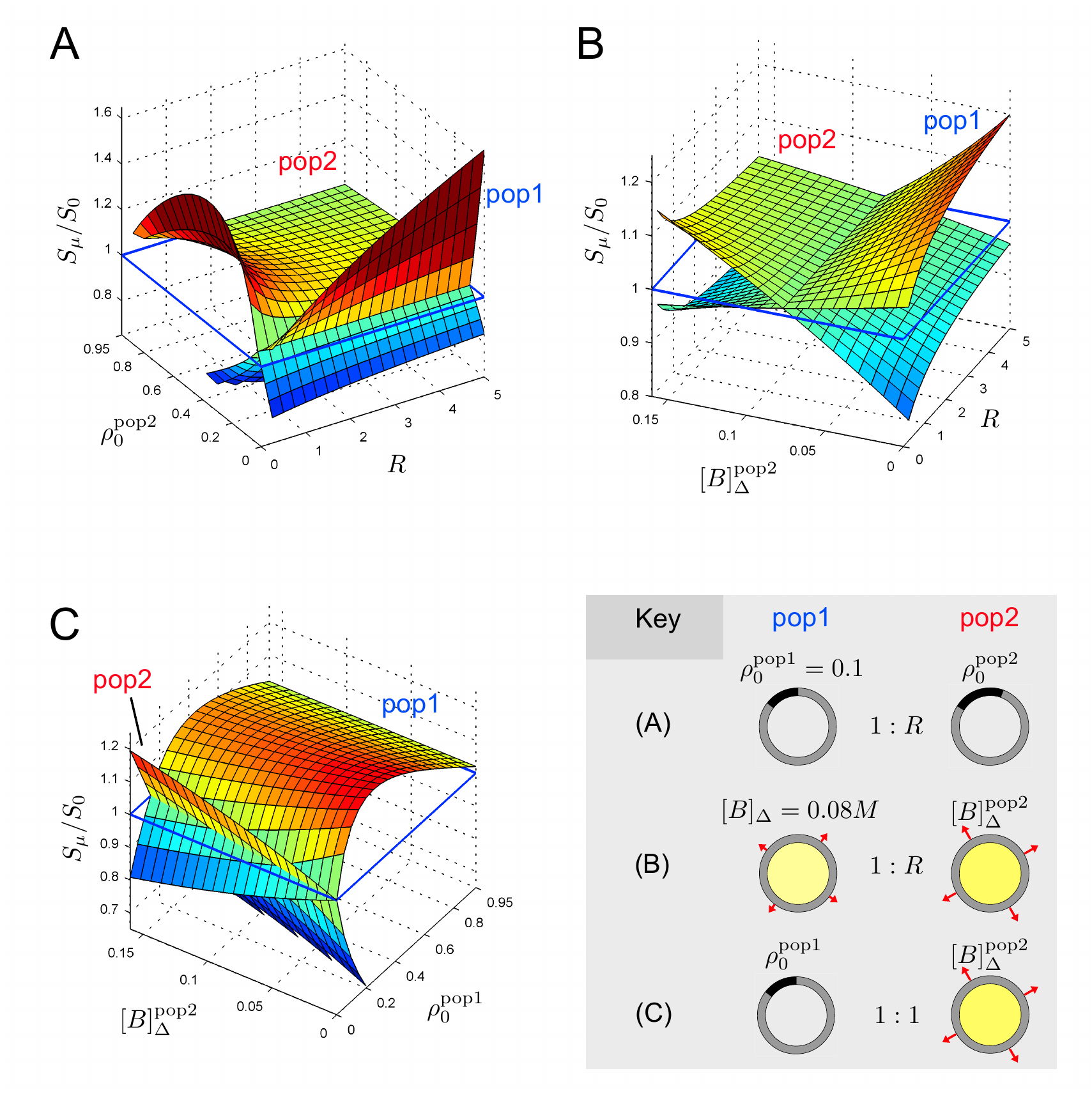}
\end{center}
\caption{
{\bf Wider exploration of three different vesicle competition scenarios.} Relative surface growths of two vesicle populations is explored in a broader context for three different competition scenarios detailed by the key. \textbf{A} Phospholipid competition. Population 1, a fixed population of vesicles with initial DOPA phospholipid fraction $\rho_0^{pop1}=0.1$, is mixed $1:R$ with population 2, whose vesicles have initial DOPA fraction $\rho_0^{pop2}$.  \textbf{B} Osmotic competition. Population 1, a fixed population of vesicles swelled by residual buffer $[B]_\Delta=0.08M$, is mixed $1:R$ with population 2, whose vesicles are swelled by residual buffer $[B]_\Delta^{pop2}$. \textbf{C} Phospholipid versus osmotically-driven competition. Vesicles with initial DOPA fraction $\rho_0^{pop1}$ are mixed $1:1$ with pure oleate vesicles swelled by residual buffer $[B]_\Delta^{pop2}$. In all cases, for DOPA laden vesicles, the direct effect is maximally enabled ($d=1$). Blue lines on plots highlight when the relative surface growth is 1.}
\label{fig:competition3d}
\end{figure*}

Finally, we were able to explore more widely some of the the parameter space for phospholipid-driven and osmotically-driven competition, using our model to make some predictions. Figure \ref{fig:competition3d}A shows the stoichiometry results of phospholipid-driven competition in this wider context. A population of DOPA:OA ($\rho=0.1$) vesicles is mixed with a second population, but the phospholipid content of the second population, as well as the mixing ratio $R$, are varied. Taking a slice through the surface labelled `pop1' when $\rho^\text{pop2}_0 = 0$ shows the result reported in Fig. \ref{fig:experimental_comparison}C as the red line. Interestingly, this figure predicts that the absolute growth of the second population of vesicles will be maximal, when their phospholipid fraction is around 50\%, and will decline again towards no overall growth as the phospholipid fraction approaches 100\%. Figure \ref{fig:competition3d}B explores the stoichiometry of osmotically-driven competition in a similar way to phospholipid-driven competition. A fixed population of swelled vesicles is mixed with a second population, and the trapped residual buffer inside vesicles in the second population, as well as the mixing ratio $R$ to the second population, are varied. To conclude these predictions, Fig. \ref{fig:competition3d}C shows the effects of osmotically-driven versus phospholipid-driven competition, still a completely unreported scenario in the experimental literature, whereby a population of swelled pure oleate vesicles competes for lipid with a population of DOPA:OA vesicles.

\noindent
\begin{table*}
\begin{tabular}{|l|l|l|l|}
\hline 
Parameter & Description & Value & Unit \\
\hline
$S_\mu^0$ & \scriptsize {Surface area of 100nm spherical vesicle} & \scriptsize {$3.142\times10^{4}$} & {\scriptsize {[}$nm^2${]}}\\ 
$\Omega^0$ & \scriptsize {Volume of 100nm spherical vesicle} & \scriptsize {$5.236\times10^{5}$} & {\scriptsize {[}$nm^3${]}}\\
$[B]$ & \scriptsize {Buffer concentration} & \scriptsize {$0.2$} & {\scriptsize {[}$M${]}}\\
$k_\text{out}$ & \scriptsize {OA lipid release constant} & \scriptsize {$7.6\times10^{-2}$} & {\scriptsize {[}$s^{-1}${]}}\\
$k_\text{in}$ & \scriptsize {OA lipid uptake constant} & \scriptsize {$7.6\times10^{3}$} & {\scriptsize {[}$s{}^{-1}$$M^{-1}$$nm^{-2}${]}}\\
$\alpha_L$ & \scriptsize {OA lipid head area$^{(1)}$} & \scriptsize {0.3} & {\scriptsize {[}$nm^{2}${]}}\\
$\alpha_P$ & \scriptsize {DOPA lipid head area$^{(1)}$} & \scriptsize {0.7} & {\scriptsize {[}$nm^{2}${]}}\\
$\epsilon$ & \scriptsize {OA vesicle burst tolerance} & \scriptsize {$0.21$} & {}\\
$[L]_\text{eq}^\text{OA}$ & \scriptsize {100nm OA vesicle, OA monomer equilibrium concentration} & \scriptsize {$6.667\times10^{-5}$} & {\scriptsize {$[M]$}}\\
$[L]_\text{eq}^\text{DOPA:OA}$ & \scriptsize {100nm DOPA:OA vesicle, OA monomer equilibrium concentration} & \scriptsize {$5.294\times10^{-5}$} & {\scriptsize {$[M]$}}\\
$N^\text{OA}$ & \scriptsize {100nm OA vesicle aggregation number} & \scriptsize {209439} & {\scriptsize {total lipids}}\\
$N^\text{DOPA:OA}$ & \scriptsize {100nm DOPA:OA vesicle aggregation number} & \scriptsize {184799} & {\scriptsize {total lipids}}\\
$\Omega_\text{stoi}$ & \scriptsize {Competition volume unit for stoichiometric calculations} & \scriptsize {$3.478\times10^{-13}$} & {\scriptsize {[}$dm^{3}${]}}\\
$\Omega_\text{dyn}$ & \scriptsize {Competition volume unit for dynamics simulations} & \scriptsize {$6.956\times10^{-15}$} & {\scriptsize {[}$dm^{3}${]}}\\
$C_0$ & \scriptsize {Mix concentration unit} & \scriptsize {0.001} & {\scriptsize {[}$M${]}}\\
$[B]_\Delta^\text{max}$ & \scriptsize {Residual buffer concentration inside maximally swelled OA vesicles$^{(2)}$} & \scriptsize {$0.16$} & {\scriptsize {$[M]$}}\\
\hline
\end{tabular}
\caption{
{\bf Vesicle competition model parameters.}}
\label{table:parameters}
\end{table*}

%
%

\section{Discussion}

In this work, we have presented a theoretical model of the transfer kinetics of lipid molecules between vesicles. We have shown that reasonable rate equations chosen for simple lipid uptake and release allow to reproduce fairly well data from controlled laboratory experiments on phospholipid-driven competition and osmotically-driven competition. Furthermore, we have been able to predict the outcome of several yet-to-be-performed experiments. Thus, it is time to recapitulate, considering possible limitations of our approach, clarifying several points that remain open, and giving a more general perspective on the problem addressed.

The main assumption we made when modelling phospholipid-driven competition is that the membrane phospholipid fraction is approximately stationary with respect to the timescale of simple lipid transfer between the supramolecular structure (i.e., the membrane bilayer) and the aqueous solution (both inwards and outwards)\footnote{ Likewise, the assumption we make with osmotically-driven competition is that the residual buffer inside the vesicles permeates very slowly through the bilayer membrane.}. In reality, the off-rate of a lipid molecule from a bilayer is inversely proportional to the number of carbon atoms in the acyl chain of the lipid concerned, and phospholipids do have a small non-zero transfer rate (with a half time from hours to days \cite{budinSzostak2011}). If phospholipid transfer was included in our model, the equilibrium reached in the limit of time would always be that of a completely homogeneous population. This is because the $P$ phospholipid would redistribute amongst the vesicles until all were equilibrated with the same phospholipid monomer concentration in solution $[P]_\text{eq}$, which is trivially when all vesicles have the same number of membrane phospholipids $P_\mu$. With no remaining asymmetries in $P_\mu$ to drive competition, all vesicles would finish with the same number of simple lipids $L$. The appearance and then disappearance of competition would follow the same type of dynamics as those experimentally reported for nervonic acid \cite[Fig. 4D, therein]{budinSzostak2011} which redistributes between vesicles (simulation results not shown). However, if vesicles contained a metabolism which synthesised phospholipid, then lasting $P_\mu$ asymmetries between vesicles could be continually maintained as steady states, in spite of the exchanging $P$ fraction. The results of this study can be interpreted as the competition advantage bestowed upon a vesicle by a membrane phospholipid fraction \emph{given that this fraction is somehow maintained as constant}. Not explicitly modelling phospholipid synthesis processes grants a simplified lipid scenario (i.e. a materially-closed system which subsequently settles to equilibrium) where some analysis can be carried out.

The next point that deserves discussion is the role of the \emph{direct effect} in driving the phospholipid-driven competition simulations performed with our model. A first observation to make is that even when the direct effect is disabled ($d=0$), the remaining indirect effect can account for the majority of the vesicle surface growth observed experimentally in phospholipid-driven competition (blue lines, Figs \ref{fig:experimental_comparison}C and \ref{fig:experimental_comparison}D). Thus, whilst a direct effect could improve the fit to experimental results, we should conclude from our treatment of the problem and the results obtained, that the indirect effect is the main mechanism driving vesicle growth dynamics. A second observation is that, as stated in the Results in reference to Fig. \ref{fig:tipping_points}B, the exact contribution of the direct effect depends on the specific details of the competition scenario. In the case of the latter figure, the lipid release multiplier function \textbf{r} linearly decreases the simple lipid off-rate with increasing phospholipid content, but this `context dependence' should also be true for different choices of function \textbf{r}.

One curiosity in the results (both in vitro and in silico) is how DOPA:OA ($\rho=0.1$) vesicles grow continually as more OA vesicles are added (Fig. \ref{fig:experimental_comparison}C). This is unintuitive, since the growth of the DOPA:OA vesicles should imply a dilution of their phospholipid content, which would seemingly reduce the indirect and direct effects, thus giving a negative feedback to eventually curb the DOPA:OA growth profile. The reason why our model reproduces this continuous growth result has to do with the mathematics underlying the kinetic modelling. In the limit of infinite $L_\mu$ lipids in the membranes of our model DOPA:OA vesicles, the inside/outside lipid concentration required to sustain them at equilibrium  (given by function $f$, (\ref{eq:function-f})) tends to \emph{but crucially never actually reaches} the CVC concentration of pure oleic acid:

\begin{equation}
\underset{L_{\mu}\rightarrow\infty}{lim} f = \frac{2k_\text{out}}{k_\text{in} \alpha_L} = [L]_\text{eq}^{OA}
\label{eq:function-f-limit}
\end{equation}

This is true, even if a model DOPA:OA vesicle contains just one single phospholipid in the membrane. Now, as more OA vesicles are mixed with the DOPA:OA vesicles, the population becomes increasingly dominated by OA vesicles and the lipid monomer concentration in the environment subsequently rises toward $[L]_\text{eq}^{OA}$. As this happens, (\ref{eq:function-f-limit}) implies that the DOPA:OA vesicles will be absorbing more and more $L$ lipids, in order to grow to a size in equilibrium with the external lipid monomer concentration. The DOPA:OA growth is thus halted only by the number of lipids in the system being limited to $L_t$. In our kinetics model, this continuous growth happens with or without the direct effect present. The direct effect can drive larger growths when vesicles are small, but as surface area increases and the direct effect diminishes, it is the indirect effect which persists and continues to drive growth as more OA vesicles are added.

A final point worth highlighting is that when the lipid uptake function $\textbf{u}$ given in (\ref{eq:function-u}) is not conditional, as we assumed, but simply

\begin{equation}
\textbf{u}({\Phi})= exp\left(\frac{1}{\Phi}-1\right)
\label{eq:function-u-continuous-only}
\end{equation}

for all membrane states (which denotes that even flaccid vesicles have differential rates of lipid uptake), then, quite interestingly, the continuous DOPA:OA growth effect cannot be reproduced. In this case, it can be shown that

\begin{equation}
\underset{L_{\mu}\rightarrow\infty}{lim} f = \frac{2k_\text{out}}{k_\text{in} \alpha_L}\cdot exp(1) > [L]_\text{eq}^{OA}
\label{eq:function-f-limit2}
\end{equation}

meaning that the DOPA:OA vesicles do not show the same continued growth as the lipid monomer concentration in the environment rises toward $[L]_\text{eq}^{OA}$. Rather, the DOPA:OA have much slower growth, and they even have a finite stable size when the outside lipid monomer concentration is exactly $[L]_\text{eq}^{OA}$. Thus, to best reproduce experimental outcomes, a crucial part of our lipid uptake kinetics was to accelerate lipid uptake \emph{only} in osmotically stressed vesicle states, not in flaccid ones.

%
%

Understanding in full detail the dynamics of these colloidal systems is certainly not an easy task. In any case, we consider this work just as a step further in the development of semi-realistic, coarse-grained descriptions of phenomena that, in reality, are extremely complex. Self-assembly processes involving heterogeneous component mixtures and the formation of dynamic supramolecular structures that could hypothetically lead to biologically relevant forms of material organization, like protocells \cite{mouritsen2005,rasmussen2009}, constitute a tremendous challenge, indeed, both for experimental and theoretical `systems chemistry' research \cite{ruizMirazo2014} and for synthetic biology \cite{soleReproductionComputation2007,sole2009}. In particular, the connection between basic metabolic reaction networks and membrane dynamics (including stationary growth and division cycles \cite{mavelliRuizMirazo2013}) needs to be explored much more extensively, since it is one of the key aspects to establish a plausible route from physics and chemistry towards biological phenomenology.

%
%

\section*{Acknowledgments}

R.S. and B.S.-E. acknowledge support from the Botin Foundation and by the Santa Fe Institute. K.R.-M. acknowledges support from the Basque Government (Grant IT 590-13), Spanish Ministry of Science (MINECO Grant FFI2011-25665), COST Action CM 1304 (Emergence and Evolution of Complex Chemical Systems). F.M. acknowledges support from MIUR (PRIN 2010/11 2010BJ23MN\_003).

%
%

\bibliographystyle{apalike}
\bibliography{refs}

\end{document}